\documentstyle[aps,multicol,epsfig]{revtex}
\begin{document}
\title{Positioning  and clock synchronization through entanglement}
\author{Vittorio Giovannetti, Seth Lloyd$^*$, and Lorenzo Maccone}
\address{Massachusetts Institute of Technology,\\ Research
Laboratory of Electronics, 50 Vassar St.\\ $^*$Department of
Mechanical Engineering MIT 3-160,\\ Cambridge, MA 02139, USA.}
\date{\today}
\maketitle

\begin{multicols}{2}
\begin{quote}
{\bf Abstract.} A method is proposed to employ entangled and squeezed
light for determining the position of a party and for synchronizing
distant clocks. An accuracy gain over analogous protocols that employ
classical resources is demonstrated and a quantum-cryptographic
positioning application is given, which allows only trusted parties to
learn the position of whatever must be localized. The presence of a
lossy channel and imperfect photodetection is considered. The
advantages in using partially entangled states is discussed.
\end{quote}

From the realm of thought experiments, quantum entanglement has
recently become exploitable for various applications and almost ready
for technological implementations in fields such as quantum
cryptography {\cite{bb84}}. Other applications for entanglement and
squeezing have been proposed in fields such as interferometric
measurements {\cite{interf}}, frequency measurements 
{\cite{wine,plenio}},
lithography {\cite{lith}}, algorithms {\cite{grover}}, {\it etc.} In
this paper a recent proposal {\cite{paper}} to exploit entanglement
and squeezing to enhance the accuracy of position measurements and
clock synchronization is thoroughly analyzed.

In Sect. {\ref{s:propos}} the proposal of {\cite{paper}} is briefly
reviewed and the notation that will be employed is presented. The
positioning protocol is derived and its main features are described.
In Subsect. {\ref{pippo1}} it is shown that our protocol gives an
enhancement in accuracy by comparing it with classical procedures that
employ analogous resources. In Subsect. {\ref{pippo2}} its use in a
cryptographic context is addressed.  In particular, two different
crypto-positioning schemes are derived that prevent non-trusted
parties to recover the position of what must be localized. The first
is essentially a classical protocol, but allows an accuracy
enhancement of the localization procedure over the unentangled
case. The second is a quantum crypto-positioning scheme derived from
the quantum cryptographic BB84 protocol {\cite{bb84}}.  In
Sect. \ref{s:loss} the analysis of the protocol is given in the
presence of loss, by considering the possibility that some photons are
lost through dissipative processes during their travel or at the
detection stage.  The loss of a single photon in the maximally
entangled state makes the resulting state completely useless. On the
other hand, the loss of a photon in the unentangled case is not so
dramatic since information on the time of arrival of the pulse may
still be obtained by measuring the times of arrival of the remaining
photons. However, by comparing the time of arrival information that
can be obtained in the two cases, one sees that one still does better
by using entangled states in a wide range of cases. The robustness to
loss stems from the fact that the accuracy gain obtained through
entanglement is high enough to beat the classical (unentangled case)
accuracy even when some of the time of arrival data must be discarded.
In Sect. {\ref{s:nonideal}}, the assumption of using maximally
entangled states is relaxed.  There is a trade-off between the degree
of entanglement (or the accuracy gain) and the robustness to noise, as
was also shown in {\cite{plenio}}.  A higher robustness against loss
ensues by decreasing the degree of entanglement, at the cost of
reducing the accuracy gain achievable. Given the loss of the available
channel, one will have to optimize the states to be employed. A scheme
which is analogous to fault tolerant quantum computation is
presented. It is possible to protect, at least partially, the
entanglement from the loss by devising entangled states where the loss
of one or more photons allows some information to be retained from the
photons which do arrive. An example of such states is derived in
detail.

\section{Positioning through entanglement}\label{s:propos}
In this section a brief review of the method proposed in
{\cite{paper}} is given. The positioning problem is defined and the
formalism that will be used in the rest of the paper is laid out. In
Subsect. {\ref{pippo1}} the enhancement in the positioning obtained by
using entangled-squeezed states is given and analyzed, by comparing it
to what one would obtain with classical states of equal spectral
characteristics. Subsect. {\ref{pippo2}} is devoted to discussing the
use of the proposed protocol in a crypto-positioning context.

For the sake of simplicity, consider the one-dimensional case in which
one party (say Alice) wants to measure her distance from the
detectors' position $x$ by sending a light pulse to each of the $M$
detectors which are placed in a known position. Alice's position can
be obtained by measuring the pulses travel time average $\langle
t\rangle$ divided by the pulses velocity.  Given the spectral
characteristics of each pulse, its time of arrival $t_i$ will have an
intrinsic indetermination. The unsurpassable limit for classical
measurements is given by the shot noise limit: one must at least
measure a single photon. The accuracy of the distance $x$ measurement
depends on the variance $\Delta t^2$ of the statistical variable
average time of arrival $\langle t\rangle$. This variance can be
related to the intrinsic accuracy $\Delta\tau^2$ achievable on the
measurement of the single photon time of arrival, which, in turn, will
ultimately depend on the photon's bandwidth.

The formalism is now introduced. The probability to detect a photon at
time $t$ and at position $x$ in an ideal photodetector with infinite
time resolution is given by the Glauber-Mandel formula
{\cite{glauber,mandel}}
\begin{equation} 
P(t) \propto
\left\langle E^{(-)}(t-x/c)E^{(+)}(t-x/c)\right\rangle \;,
\label{probm1}
\end{equation}
where the ensemble average is the expectation on the quantum state of
the radiation.  All actual photodetectors are of course non-ideal, but
the fundamental limit to the error introduced by the non-ideal
features of photodetectors is given by the bandwidth of the
photodetector rather than the bandwidth of the detected photon
\cite{PERES}.  In addition, this error can in principle be made as
small as desired by devoting more resources (of energy, power, etc.) to
the photodetection process.  In Eq. (\ref{probm1}), the signal field
at time $t$ is given by
\begin{eqnarray}E_i^{(-)}(t)\equiv
\int d\omega\  a_i^\dag(\omega)\; e^{i\omega t}
\;;\quad E_i^{(+)}\equiv\left(E_i^{(-)}\right)^\dag\;\label{campocont},
\end{eqnarray}
where $a_i(\omega)$ is the field annihilator of a quantum of frequency
$\omega$ at the $i$-th detector position. In the continuous Fock space
formalism {\cite{schweb}} of Eq. (\ref{campocont}), the field
annihilation operator is not dimensionless and satisfies the
commutation relation
\begin{eqnarray}
[a_i(\omega),a^\dag_j({\omega'})]=\delta_{ij}\delta(\omega-\omega')
\;\label{commutatori},
\end{eqnarray}
where the Kronecker delta accounts for the independence of the
channels. The electromagnetic field has been quantized so that
$E^{(-)}E^{(+)}$ is given in units of photons per second.  For $M$
different communication channels, each of which may receive more than
one photon, Eq. (\ref{probm1}) generalizes to
\begin{equation} 
P_M(t_{i,k};N_i) \propto
\left\langle : \prod_{i=1}^{M} \prod_{k=1}^{N_{i}} 
E_i^{(-)}(t_{i,k})E_i^{(+)}(t_{i,k}) : 
\right\rangle \;,
\label{probmisn}
\end{equation}
where $t_{i,k}$ is the time of arrival of the $k$-th photon in the
$i$-th channel, $N_i$ is the number of photons detected in the $i$-th
channel, and the detection time is shifted by the detector's position
$x_i$: $t_{i,k}\to t_{i,k}+x_i/c$. The probability $P_M(t_{i,k};N_i)$
must be normalized so that, when integrated over all the arrival times
$t_{i,k}$, it gives the probability of detecting $N_i$ photons in the
$i$-th channel.  In the case of unit quantum efficiency $\eta=1$ (when
no photons are lost through dissipative processes), this is also the
probability of having $N_i$ photons in the channel. In the case
$\eta<1$ this is not true anymore, because there is a probability
$1-\eta$ that a photon will be lost in the channel or at the
photodetection stage. A detailed analysis of this case is given in
Sect. \ref{s:loss}. In the cases of coherent states and of states with
definite number of photons that will be considered here, this choice
of normalization allows to use the formula (\ref{probmisn}) instead of
the more complicated conditional joint probability (see
{\cite{mandel}} Chap. 14.8) of measuring {\it only} $N_i$ photons at
times $t_{i,k}$ and no more in each of the $M$ channels.

Consider the situation where all the detectors are placed at the same
position $x$. The probability $P_M(t_{i,k};N_i)$ of
Eq. (\ref{probmisn}) contains all the timing information relative to
the transmitted pulses sent by Alice. In particular, the average time
of arrival $\langle t\rangle$ needed for the position measurement can
be obtained by taking the average of the quantity
\begin{eqnarray} T\equiv\frac{1}{M}\sum_{i=1}^M\frac
1{N_i}\sum_{k=1}^{N_{i}}t_{i,k}
\;\label{deftgrande}
\end{eqnarray}
over the probability $P_M(t_{i,k};N_i)$, namely
\begin{eqnarray}
\langle t\rangle=\sum_{N_i}\int dt_{i,k}\;P_M(t_{i,k};N_i)\;T
\;\label{tmedio},
\end{eqnarray}
where the sum is performed on the values of $N_i$ for all $i$ and the
integration is performed on all the $t_{i,k}$.  The statistical error
in determining $\langle t\rangle$ from the measurement results is
given by the variance of $T$. This variance is dependent on the shape
of the probability $P_M$, which in turn depends on the quantum state
of the impinging light pulses, through Eq. (\ref{probmisn}).

\subsection{Quantum enhancement}\label{pippo1}
Consider first the case of unit quantum efficiency $\eta=1$, where no
photons are lost. The $M$ coherent pulses a ``classical'' Alice would
send to the reference detectors are described by a state of the
radiation field of the form
\begin{eqnarray} |\Psi\rangle_{cl}=\bigotimes_{i=1}^M
\bigotimes_\omega
\left|\alpha\left[\phi(\omega)\sqrt{N}\right]\right\rangle_i
\;\label{statoclassico},
\end{eqnarray}
where $\omega$ is the pulses carrier frequency, $\phi(\omega)$ is
their spectral function,
$\left|\alpha[\lambda(\omega)]\right\rangle_i$ is a coherent state of
frequency $\omega$ and amplitude $\lambda(\omega)$ directed towards
the $i$-th detector, and $N$ is the mean number of photons in each
pulse. The pulse spectrum $|\phi(\omega)|^2$ has been normalized so
that $\int d\omega|\phi(\omega)|^2=1$.  Upon calculating the ensemble
average of Eq. (\ref{probmisn}) with the state $|\Psi\rangle_{cl}$
using the property
\begin{eqnarray}
a(\omega')\bigotimes_\omega\left|\alpha[\lambda(\omega)]\right\rangle
=\lambda({\omega'})\bigotimes_\omega\left|\alpha[\lambda(\omega)]\right
\rangle
\;\label{derivaz},
\end{eqnarray}
one obtains the probability density
\begin{eqnarray} P_M(t_{i,k};N_i) \propto \prod_{i=1}^M 
\prod_{k=1}^{N_{i}} |g(t_{i,k})|^2
\;\label{classicp},
\end{eqnarray}
where $g(t)$ is the Fourier transform of the spectral function
$\phi(\omega)$: \begin{eqnarray} g(t)=\frac 1{\sqrt{2\pi}}\int
d\omega\;\phi(\omega)\;e^{-i\omega t}
\;\label{defdig}.
\end{eqnarray}
Notice that the probability $P_M$ factorizes, since in the
classical state all the photons are independent. The quantity
$|g(t_{i,k})|^2$ is the probability that the $k$-th photon is received
on the $i$-th channel at time $t_{i,k}$. Define $\Delta\tau^2$ as the
variance of $|g(t_{i,k})|^2$ (which is independent on $i$ and $k$ since
all the photons have the same spectrum). From Eq. (\ref{classicp}) it
follows that the statistical error relative to the mean time of
arrival $\langle t\rangle$ is
\begin{eqnarray} 
&&\Delta t \gtrsim \frac {\Delta\tau}{\sqrt{MN}}\;,
\label{classictime}
\end{eqnarray}
with approximate equality for $N\gg1$. 

Now compare this result with the one obtained from a quantum state
which combines entanglement and photon number squeezing.  Define
number squeezed state of frequency $\omega$ the state
$|N_\omega\rangle$ in which all modes are in the vacuum state, except
for the mode at frequency $\omega$ which is populated by exactly $N$
photons. The entangled-squeezed state that allows to achieve the most
enhancement over the classical case is given by \begin{eqnarray}
|\Psi\rangle_{NM}=\int d\omega
\;\phi(\omega) |N_{\omega}\rangle_1\cdots |N_{\omega}\rangle_M
\;\label{defpsinm}.
\end{eqnarray}
By choosing the same spectral function $\phi(\omega)$ of the state
(\ref{statoclassico}), the spectral characteristics of each of the
channels of the state $|\Psi\rangle_{NM}$ (obtained by tracing
$|\Psi\rangle_{NM}$ over all the other channels) is the same as the
classical state.  Notice that $|\Psi\rangle_{NM}$ is a frequency
maximally entangled state: a measurement of the frequency of a single
one of its photons will have a random outcome weighted by the
probability $|\phi(\omega)|^2$, but will determine the frequency of
all the other photons. Since the number of photons in each channel is
fixed ($N$) and no photons are lost ($\eta=1$), then the probability
$P_M(t_{i,k};N_i)$ is null for $N_i\neq N$, thanks to its
normalization discussed previously. For $N_i=N$, inserting
$|\Psi\rangle_{NM}$ in Eq. (\ref{probmisn}), it follows
\begin{eqnarray}
P_M(t_{i,k};N)\propto |g(\sum_{i=1}^M\sum_{k=1}^Nt_{i,k})|^2
\;\label{pisqent},
\end{eqnarray}
where the property
$[a_i(\omega')]^N|N_\omega\rangle_j=\delta_{ij}\delta(\omega-\omega')
\sqrt{N!}|0\rangle$
was employed ($|0\rangle$ being the normalized vacuum state) and
$g(t)$ is the same of Eq. (\ref{defdig}).  Eq. (\ref{pisqent}) shows
that the entanglement in frequency translates into the bunching of the
times of arrival of the photons of different pulses: although their
individual times of arrival are random, the average $T=\frac
1{MN}\sum_{i,k}t_{i,k}$ of these times is highly peaked. Indeed, from
Eq. (\ref{pisqent}) it results that the probability distribution of
$T$ is $|g\left(MN T\right)|^2$. This immediately implies that the
average time of arrival $\langle t\rangle$ is determined to an
accuracy
\begin{eqnarray} 
\Delta t=\frac{\Delta\tau}{MN}
\;\label{valorimedi},
\end{eqnarray}
where $\Delta\tau$ is the same of Eq. (\ref{classictime}). This result
shows a $\sqrt{MN}$ accuracy improvement over the classical case
(\ref{classictime}).  The Margolus-Levitin theorem \cite{MARG} implies
that a $\sqrt{MN}$ improvement in accuracy is the best that can be
obtained {\cite{paper}}. The role of the entanglement and the role of
the squeezing in enhancing position measurements are separately
addressed in {\cite{paper}}. It is shown that the $\sqrt{M}$
enhancement derives from the entanglement between the channels and the
$\sqrt{N}$ enhancement from the number-squeezing within each channel.

Notice that when the state $|\Psi\rangle_{NM}$ is used, the results of
the single time of arrival measurement are meaningless: it is
necessary to make correlation measurements, {\it i.e.} in this case
one must consider the {\it sum} of the times of arrival of all the
photons as in the quantity $T$. This implies that the geometry of the
problem that can be solved depends on the state that can be
produced. The state $|\Psi\rangle_{NM}$, which is tailored as to give
the least indetermination in the physical quantity $T$, is appropriate
for the geometry of the case considered here, where the sum of the
pulses' time of arrival is needed. Other maximally entangled states
have to be tailored for different geometric dispositions of the
detectors {\cite{paper}}.\par

In conclusion, the suggested positioning protocol requires: 1) to
produce and deploy the maximally entangled state suited for the given
disposition of the reference points; 2) to measure the time of arrival
$t_{i,k}$ of $k$-th photon in the $i$-th reference point and 3) to
collect and compare the results in order to have the needed
correlation measurement. 

\subsection{Quantum cryptographic positioning}\label{pippo2}
The accuracy enhancement over classical protocols is not the only
reason that makes the use of quantum mechanics appealing in the
positioning problem.  In fact, one is also offered the possibility to
employ the ideas of quantum cryptography in this context.  In this
section two different crypto-positioning protocols based on our scheme
will be given. The aim is for Alice to learn her position in space
relative to Bob (located at the detectors position), without anybody
else gaining {\it any} information by intercepting neither the photons
nor the classical information Alice and Bob exchange.

The first protocol is essentially equivalent to a classical protocol
in which Alice sends Bob photons she delayed each by a random amount
of time she does not disclose. From Bob's random times of arrival she
may recover her position without anybody else (including Bob) knowing
it. In the quantum version given here, however, the accuracy for fixed
number $M$ of photons is increased over the classical version. This
protocol allows only Alice to recover her position: nobody else
(including Bob) will be able to determine where she is.  Consider for
simplicity the case of the state $|\Psi\rangle_{NM}$ with one photon
per channel ($N=1$), given by
\begin{eqnarray}
|\Psi\rangle_{en}\equiv\int d\omega
\;\phi(\omega)\;|\omega\rangle_1\cdots|\omega\rangle_M
\;\label{freqmes},
\end{eqnarray}
where $|\omega\rangle\equiv|1_\omega\rangle$. The extension to the
general case is straightforward. This protocol is simply implemented
by allowing Alice to detect the time of arrival of the photons in one
of the $M$ channels. She will send to Bob only the rest $M-1$
photons. When Bob receives and measures them, he will use a public
channel to broadcast the measurement result to Alice. As will be shown
in Sect. {\ref{s:loss}}, the loss of a single photon results in not
being able to recover {\it any} information on Alice's position. Thus
if an eavesdropper was to intercept the photons Alice sends Bob (the
eavesdropper needn't even bother: he only has to wait for Bob's
broadcast) he would obtain no information. Alice, on the other hand,
simply has to add the random times of arrival that Bob tells her to
the one she herself has measured. This allows her to find her
position, with an uncertainty $\Delta t=\Delta\tau/(M-1)$, since she
only used $M-1$ photons for the positioning.

The second protocol allows both Alice and Bob to recover their
distance without anybody else discovering it. This protocol is
analogous to the quantum cryptographic key exchange BB84
{\cite{bb84}}.  Alice and Bob share $r$ copies of the state
$|\Psi\rangle_{en}$ of which, as before, Alice retains one photon and
sends Bob the remaining $M-1$. For each of the $r$ copies Alice and
Bob choose randomly (and independently) to measure either the
frequency or the time of arrival of {\it all} the photons. After that
they compare which of the two ``observables'' they used on each of the
$r$ copies they exchanged: they discard all the cases in which the two
observables do not match, namely Alice measured the frequency and Bob
the time of arrival or {\it viceversa}. For all the cases in which
both of them measured the frequency, they broadcast the measurement
results. Since the state is maximally entangled in frequency, their
measurement outcomes (though random) must agree. If this is not the
case, they know that there is an eavesdropper which is ruining the
states that are transiting between them. If all the frequency
measurement outcomes do agree, they can be confident that no one is
measuring the photon time of transit in the channel. Once they
verified that no eavesdropper was present, Alice can broadcast the
measurement results for half of the copies in which they both measured
the time of arrival and Bob can broadcast the measurement results of
the other half. From the information they exchange, which is utterly
useless for anybody else, both Alice and Bob may recover Alice's
position.  Of course an eavesdropper might be measuring the frequency
of the exchanged photons without being detected, but this will not
give him any information on Alice's position: he may only succeed in
ruining Alice and Bob's exchange.

Notice that it is possible to modify this second protocol to include
more complicated scenarios, such as the case in which also other
trusted persons may be allowed to learn Alice's position, or (by
suitably tailoring the entanglement of the exchanged pulses) the case
in which some of the trusted persons may learn Alice's position {\it
only} when they meet and exchange their data, or the case in which
Alice herself is not allowed to discover her own position, {\it etc.}

Finally, it is worth to notice that an implementation of the
crypto-positioning schemes described here can be achieved with the
state $|\Psi\rangle_{en}$ for $M=2$ the practical realization of which
has been recently proposed in {\cite{shapiro}}.

\section{Loss analysis in the ideal case}\label{s:loss}
In this section the problem of the loss is addressed. The loss of a
single photon from a maximally entangled state (such as
$|\Psi\rangle_{MN}$) makes it completely useless for positioning,
since the information is encoded in the entanglement and not on the
single photons. On the other hand, the loss of a single photon from a
``classical'' state (such as $|\Psi\rangle_{cl}$) allows still to
recover information on the time of arrival of the remaining
photons. Nonetheless, it will be shown that the the gain in accuracy
obtained by using entangled photons {\it vs.} unentangled is quite
robust against the loss.  In the first subsection the conditions on
the channel quantum efficiency that is necessary to obtain an
enhancement in the accuracy is derived.  First a simple argument is
given, then a more rigorous approach is discussed.  In
Subsect. {\ref{s:pluto2}} the effect of the loss on the state is
studied in the density matrix formalism.

\subsection{Condition on the quantum efficiency}\label{s:pluto1}
\begin{figure}[hbt]
\begin{center}\epsfxsize=.6
\hsize\leavevmode\epsffile{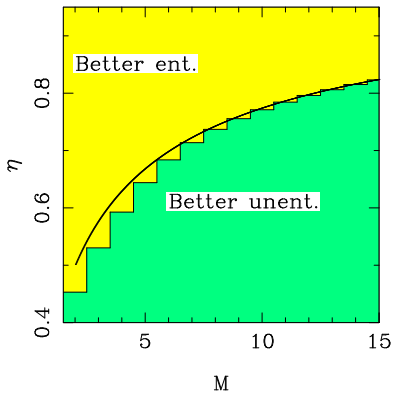} 
\end{center}
\caption{ Graph showing which values of quantum efficiency $\eta$ are
needed to achieve an accuracy increase with the entangled state
$|\Psi\rangle_{en}$ of $M$ photons over the unentangled state
$|\Psi\rangle_{un}$ of $M$ photons. The higher region is where a
better accuracy may be obtained using $|\Psi\rangle_{en}$ and the
lower region is where a better accuracy is obtained through
$|\Psi\rangle_{un}$. The continuous line graphs the condition
(\ref{condiz1}). The histogram is obtained by the more rigorous
analysis of Eq. (\ref{condizok}). The two conditions coincide for
$M\gg 1$. }
\label{f:grafico}\end{figure}

One can understand the robustness to loss from the following intuitive
explanation (the rigorous derivation is given in detail later). For
simplicity, consider the case of one photon per channel ($N=1$),
comparing the entangled state $|\Psi\rangle_{en}$ given in
Eq. (\ref{freqmes}) with its unentangled analogous ({\it i.e.} with
one photon per channel) given by
\begin{eqnarray} |\Psi\rangle_{un}=\bigotimes_{i=1}^M\int
d\omega_i\;\phi({\omega_i})\;|\omega_i\rangle_i
\;\label{frequnent},
\end{eqnarray}
which describes $M$ uncorrelated single photon pulses each with the
same spectral function $\phi(\omega)$ of (\ref{freqmes}).  Given the
channels' quantum efficiency $\eta$ (namely $1-\eta$ is the
probability that one photon is lost), then the probability that all
$M$ photons reach Alice is given by $\eta^M$.  Repeating $r\gg 1$
times the whole experiment, a total number $r\; M$ of photons is
sent. In average only a fraction $\eta^M$ of the experimental runs
will not lose any photon.  If Alice is employing the entangled states
$|\Psi\rangle_{en}$ of Eq. (\ref{freqmes}) ({\it i.e.} the state
$|\Psi\rangle_{NM}$ with $N=1$) to evaluate the mean time of arrival
$\langle t\rangle$, she must only use the data obtained from the
experimental runs where all the $M$ photons of the state reach the
detectors. As will be shown, the other cases in which some of the
photons are lost are useless. The evaluation of the time of arrival
accuracy obtained from the $r$ experimental runs through
Eq. (\ref{valorimedi}) will then be
\begin{eqnarray}
\Delta t{(r)}=\frac{\Delta\tau}{M\sqrt{r\eta^M}}
\;\label{accu1},
\end{eqnarray}
where the factor $1/\sqrt{r\eta^M}$ stems from the statistical
independence of different experimental runs. On the other hand, if
Alice employs $r$ copies of the unentangled $M$ photon state
$|\Psi\rangle_{un}$ defined in (\ref{frequnent}), all of the $\eta\:
rM$ photons that in average reach the detectors may be employed to
evaluate the time of arrival with an accuracy
\begin{eqnarray}
\Delta t{(r)}\gtrsim\frac{\Delta\tau}{\sqrt{\eta rM}}
\;\label{accu2},
\end{eqnarray}
where the equality holds for $rM\gg 1$.  The condition for achieving a
greater accuracy through the state $|\Psi\rangle_{en}$ than through
$|\Psi\rangle_{un}$ is given by
\begin{eqnarray}
\frac{\Delta\tau}{\sqrt{\eta rM}}>\frac{\Delta\tau}{M\sqrt{r\eta^M}} 
\quad\Longrightarrow\quad \eta>\left(\frac 1M\right)^{\frac 1{M-1}}
\;\label{condiz1}.
\end{eqnarray}
This condition is shown in Fig. {\ref{f:grafico}}. It is evident that
relatively low values of quantum efficiency $\eta$ are sufficient for
obtaining the accuracy increase feature also for high numbers of
entangled photons.

\begin{figure}[hbt]
\begin{center}\epsfxsize=.75
\hsize\leavevmode\epsffile{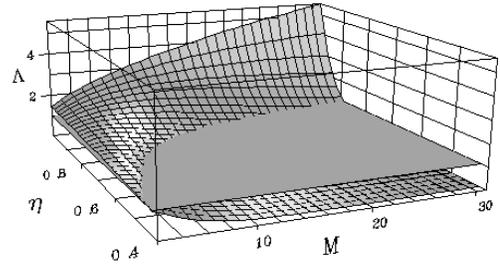} 
\end{center}
\caption{Three dimensional graph depicting the gain in accuracy
$\Lambda(M,\eta)$ {\it vs.} the number of photons $M$ and the quantum
efficiency $\eta$. The horizontal plane in the figure for $\Lambda=1$
separates the regions where it is better to employ $|\Psi\rangle_{en}$
(over) and $|\Psi\rangle_{un}$ (under).  Notice the $\sqrt{M}$
dependence for $\eta=1$ which corresponds to the enhancement discussed
in Sect. {\ref{pippo1}}.}
\label{f:grafico3d}\end{figure}

The intuitive reasoning that yields the condition (\ref{condiz1}) must
be taken only as a qualitative demonstration, since Eq. (\ref{accu2})
is valid only for $rM\gg 1$. Now the rigorous condition is derived. It
turns out to be even more favorable to the entangled case, even though
only a small correction to the condition (\ref{condiz1}) is
required. Eq. (\ref{classicp}) shows that, in the case of no loss,
using an unentangled state $|\Psi\rangle_{un}$, the probability
distribution $P_M(t_1,\cdots,t_M)$ of the time of arrival of the $M$
photons is just the product of the probability distributions of the
times of arrival of the single photons $|g(t)|^2$. Thus, if each
photon has a probability $\eta$ of arriving and a probability $1-\eta$
of being lost, then the probability of retaining $m$ of the initial
$M$ photons is given by the binomial distribution
\begin{eqnarray} &&P_m(t_1,\cdots,t_m)\!\!=\!\!
\left(\matrix{M\cr
m}\right)\frac{\eta^m(1-\eta)^{M-m}}{1-(1-\eta)^M}
\prod_{i=1}^M|g(t_i)|^2
\;\label{probcl}.
\end{eqnarray}
In this case, the integral of $P_m$ over all the times of arrival
$t_1,\cdots,t_m$ is the probability of retaining $m$ of the $M$
photons, but discarding the case in which all the photons are lost, an
event that happens with probability $(1-\eta)^M$. In fact, in the
latter case no information on time of arrival is acquired and this is
the source of the renormalization factor $1/[1-(1-\eta)^M]$ in
Eq. (\ref{probcl}). In particular for $\eta=1$ Eq. (\ref{probcl})
coincides with (\ref{classicp}), namely $P_m(t_1,\cdots,t_m)=0$ for
$m\neq M$. The accuracy that may be obtained from $|\Psi\rangle_{un}$
is given by the the variance of the distribution given in
(\ref{probcl}), {\it i.e.} \begin{eqnarray}&&
\Delta t=\left[\sum_{m=1}^M
\left(\matrix{M\cr
m}\right)\frac{\eta^m(1-\eta)^{M-m}}{{m[{1-(1-\eta)^M}]}}
\right]^{\frac 12}\Delta\tau\;.
\;\label{varclass}
\end{eqnarray}
If the experiment is repeated $r\gg 1$ times, in a fraction
$1-(1-\eta)^M$ of them at least one photon is received and the
accuracy that can be reached in each of these cases is given by
(\ref{varclass}). Thus the overall accuracy for the $r$ experiments is
\begin{eqnarray}&&
\Delta t{(r)}=\left[\sum_{m=1}^M
\left(\matrix{M\cr
m}\right)\frac{\eta^m(1-\eta)^{M-m}}{m[{1-(1-\eta)^M}]^2}
\right]^{\frac 12}\frac{\Delta\tau}{\sqrt{r}}\;.
\;\label{varclass1}
\end{eqnarray}
Again, by comparing this variance with the one obtained from the
entangled case (\ref{accu1}), one finds the condition under which it
is better to use entangled states with respect to unentangled ones,
{\it i.e.}
\begin{eqnarray}
\Lambda&\equiv&
{M}\left[\sum_{m=1}^M
\left(\matrix{M\cr
m}\right)\frac{\eta^{M+m}(1-\eta)^{M-m}}
{m[{1-(1-\eta)^M}]^2}\right]^{\frac
12}> 1
\;,
\label{condizok}
\end{eqnarray}
which for $M\gg 1$ coincides with condition (\ref{condiz1}). The
condition (\ref{condizok}) is plotted in Fig. {\ref{f:grafico3d}}.

\subsection{Loss dynamical evolution}\label{s:pluto2}
In this subsection the evolution of the states introduced previously
is analyzed in the presence of loss. Also here, for simplicity, we
analyze the case $N=1$ of one photon per channel. 

\begin{figure}[hbt]
\begin{center}\epsfxsize=.3
\hsize\leavevmode\epsffile{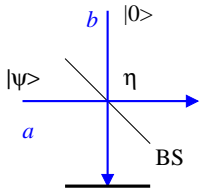} 
\end{center}
\caption{Description of a lossy channel mode through a beam splitter of
transmissivity $\eta$ equal to the channel quantum efficiency. }
\label{f:bs}\end{figure}

It can be shown {\cite{carmichael}} that a lossy channel of quantum
efficiency $\eta$ (which also takes into account the detection
efficiency) can be described by considering a perfect channel and
inserting a beam splitter of transmissivity $\eta$. The second input
port $b$ of the beam splitter is in the vacuum state $|0\rangle$ and
one output port is traced out (refer to
Fig. {\ref{f:bs}}). 
This allows to obtain the non-unitary evolution of a lossy channel. It
can be shown that, starting from the unitary evolution of the beam
splitter
\begin{eqnarray}
U=\exp\left[-\arctan\left(\sqrt{\frac{1-\eta}\eta}\right)
(ab^\dag-a^\dag b)\right]
\;\label{bsunit}
\end{eqnarray}
(where the mode definition for $a$ and $b$ is given in
Fig. {\ref{f:bs}}), one obtains the following completely positive
map for the density matrix evolution in the presence of loss
\begin{eqnarray}\varrho\longrightarrow
\varrho'=\mbox{Tr}_b\left[
U\varrho\otimes|0\rangle_b\langle 0|U^\dag\right]=\sum_{n=0}^\infty
V_n\varrho V_n^\dag
\;\label{liouv},
\end{eqnarray}
with
\begin{eqnarray}
V_n=\left(\frac{1-\eta}{\eta}\right)^{\frac n2}\frac
{a^n}{\sqrt{n!}}\eta^{\frac{a^\dag a}2}
\;\label{defvn}.
\end{eqnarray}
The case of frequency independent loss is considered. The evolution
(\ref{liouv}) must be calculated for each mode of the continuum of
modes of the entangled and unentangled states given respectively by
$|\Psi\rangle_{en}$ defined in Eq. (\ref{freqmes}) and
$|\Psi\rangle_{un}$ defined in (\ref{frequnent}).  In the case of the
density operator $\varrho_{en}=|\Psi\rangle_{en}\!\langle\Psi|$
corresponding to the state $|\Psi\rangle_{en}$, it is possible to show
\begin{eqnarray} 
&&\varrho_{en}'=\eta^M\varrho_{en}+\sum_{m=0}^{M-1}
 \eta^m(1-\eta)^{M-m}\times\nonumber\\&&\int
d\omega|\phi(\omega)|^2\Big[|\omega\rangle\langle\omega| \otimes
|0\rangle\langle 0|\otimes\cdots+|0\rangle\langle 0|\otimes\cdots
\Big]
\;\label{losssuent},
\end{eqnarray}
where $|0\rangle\langle 0|$ is the vacuum state and the term in square
brackets is the sum of all the $\left(\matrix{M\cr m}\right)$ possible
combinations of $m$ times the state $|\omega\rangle\langle\omega|$ and
$M-m$ times the vacuum $|0\rangle\langle 0|$. The interpretation of
Eq. (\ref{losssuent}) is that none of the photons is lost and the
state is unaffected with a probability $\eta^M$, and $m$ photons are
lost and the state is left in a mixture of $|\omega\rangle$ and
$|0\rangle$ with probability $\left(\matrix{M\cr
m}\right)\eta^m(1-\eta)^{M-m}$. Since the second term of the state
(\ref{losssuent}) contains only density matrices diagonal in the
$|\omega\rangle$ representation, it does not contain any information
on the time of arrival measurement. In fact, the probability $P_M$
defined in (\ref{probmisn}) gives a ``constant'' probability if
applied to the state $|\omega\rangle\langle\omega|$. Thus
post-selection measurements are needed in this case: if Alice is
expecting the state $|\Psi\rangle_{en}$, she must throw away all the
data coming from events in which she recorded less than $M$
photons. These events are useless. As shown before, the fragility to
loss is only apparent, since the accuracy gain over the unentangled
case is high enough so that it is possible to find a wide experimental
region in which the accuracy enhancement is preserved.

On the other hand, the evolution of the unentangled state
$|\Psi\rangle_{un}$ defined in Eq. (\ref{frequnent}),
$\varrho_{un}=|\Psi\rangle_{un}\!\langle\Psi|$, is given by
\begin{eqnarray}
\varrho'_{un}&=&\sum_{m=0}^M
\eta^m(1-\eta)^{M-m}\times\nonumber\\&&
\Big[\varrho_1\otimes\varrho_2\otimes\cdots+|0\rangle\langle
0|\otimes\varrho_2\otimes\cdots\Big]
\;\label{evolunen},
\end{eqnarray}
where the term in square brackets contains the sum of all possible
combinations of $m$ times the states $\varrho_i$ and $M-m$ times the
vacuum $|0\rangle\langle 0|$, and where \begin{eqnarray}
\varrho_i=\int d\omega
d\omega'\;\phi(\omega)\phi^*(\omega')|\omega\rangle_i\!\langle\omega'|
\;\label{rhoi},
\end{eqnarray}
which is a single photon wavepacket with spectral function
$\phi(\omega)$ in the $i$-th channel, {\it i.e.} the state
(\ref{frequnent}) for $M=1$.  Starting from the state in
Eq. (\ref{evolunen}) no post-selection is necessary (except the
obvious case in which Alice does not receive any photon), since all
the terms are composed of the states of the form (\ref{rhoi}) which do
retain time of arrival information.

The same analysis can be extended to the general case of the state
$|\Psi\rangle_{NM}$, showing that the loss of a single photon destroys
all the timing information.

\section{Trade-off entanglement {\it vs.} loss resistance}
\label{s:nonideal}
In this section some strategies for battling the effects of the loss
are presented. Instead of using the maximally entangled states
employed so far, one may devise strategies for using partially
entangled states which turn out to be more robust to the loss. 
The use of partially entangled states to protect entangled atomic
clocks from the effects of decoherence was noted in {\cite{plenio}}.
Here we show that partial entanglement can protect against loss
while still retaining some of the quantum enhancement. A
simple example to illustrate this is first presented and a more
sophisticated case is then analyzed in detail.

It is well known (see for example {\cite{mole}}) that when more than
two systems are entangled, variety of different effects can
occur. Hence, in order to address the relation occurring between the
degree of entanglement of a state and its loss resistance, it is
useful to start from a simple example. Consider the case of one photon
per channel ($N=1$) where the first $Q$ of the $M$ channels are
maximally entangled as the ones in the state $|\Psi\rangle_{en}$ of
Eq. (\ref{freqmes}) and the other $M-Q$ channels are unentangled as in
$|\Psi\rangle_{un}$ of Eq. (\ref{frequnent}). The parameter $Q$
characterizes the degree of entanglement of this state: bigger
values of $Q$ correspond to higher entanglement.  Consider first the
case of unit quantum efficiency. It is easy to show through
Eq. (\ref{probmisn}) that the accuracy in the determination of
$\langle t\rangle$ follows as
\begin{eqnarray}
\Delta t=\frac{\Delta\tau}{\sqrt{M}}\sqrt{\frac{{M-Q+1}}{M}}
\;\label{qui}.
\end{eqnarray}
For $Q>1$ ({\it i.e.} at least two of the $M$ channels are entangled),
the accuracy achievable is greater than the completely unentangled
case, but not as high as the completely entangled case. The loss of
performance of this state is balanced by a greater resistance to the
effects of photon losses than the maximally entangled state
$|\Psi\rangle_{en}$ for which the loss of a single photon proves
fatal. On the contrary, the loss of photons from the partially
entangled state still allows to recover information, if a suitable
post-selection is employed.  Namely one must discard all the times of
arrival of the entangled photons if one or more of them is lost, but
all the times of arrival of the unentangled photons which do arrive
can be safely retained.

This simple example shows how one can increase the resistance to loss
by reducing the entanglement, however at the cost of achieving less
accuracy enhancement. Of course much more sophisticated
configurations can be introduced for entangling multiple systems
{\cite{mole}}, in which the different systems share a different degree
of entanglement with all the other systems. It is expected that also
in the general case, a similar trade-off between the degree of
entanglement and resilience to loss holds.  Depending on the quantum
efficiency of the channel and on the degree of entanglement one is
able to produce, different strategies, involving different data
processing or post-selections, are possible. A better insight on this
may be gained by analyzing the following example, where a
multi--structured entanglement is employed.

\begin{figure}[hbt]
\begin{center}\epsfxsize=.5
\hsize\leavevmode\epsffile{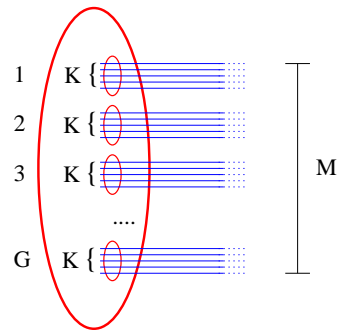} 
\end{center}
\caption{Quantum fault tolerance applied to the quantum positioning
protocol. Each of the $G$ groups of photons (which are frequency
entangled) is composed of $K$ frequency maximally entangled photons.}
\label{f:ftol}\end{figure}

A procedure analogous to fault tolerant quantum computation may be
introduced in our scheme. Consider again the simple case of one photon
in each of the $M$ channels ($N=1$). Instead of sending the maximally
entangled state $|\Psi\rangle_{en}$ of Eq. (\ref{freqmes}), Alice
sends Bob a state in which groups of $K$ photons are maximally
entangled and $G=M/K$ groups are entangled together, as depicted in
Fig. {\ref{f:ftol}}.  If no photon is lost, then one will not only be
able to use the correlations within all the groups, but also the
correlation {\it between} the groups.  In the event of a photon loss,
thanks to the structure of the entanglement employed, not all the
information will be lost as would happen when using the state
$|\Psi\rangle_{en}$. In fact, suppose that the lost photon comes from
the $j$-th group of photons: as will be shown, the only data that must
be discarded is the data relative to the $j$-th group photon times of
arrival. All the other times of arrival may be retained and
employed. The procedure can also be nested, namely each of the $G$
groups of $K$ photons may be partitioned in maximally entangled
subgroups and so on.

The state represented in Fig. {\ref{f:ftol}} is given by
\begin{eqnarray} |\Psi\rangle_G&\equiv&\int d\Omega
\;\Phi(\Omega)\;|\Omega\rangle_1|\Omega\rangle_2\cdots |\Omega\rangle_G
\;\label{omegag},
\end{eqnarray}
where
\begin{eqnarray}
|\Omega\rangle_j&\equiv&\int
d\omega\;\phi(\omega,\Omega)\;|\omega\rangle_{j1}
|\omega\rangle_{j2}\cdots|\omega\rangle_{jK}
\;\label{stato3}
\end{eqnarray}
is the state of the $j$-th group of K photons described by the one
photon frequency state $|\omega\rangle_{jl}$ for $j=1,\cdots,G$ and
$l=1,\cdots,K$. Consider for simplicity the case of Gaussian spectrum,
namely $|\Phi(\Omega)|^2$ is a Gaussian with variance $\Delta\Omega^2$
and $|\phi(\omega,\Omega)|^2$ is a Gaussian centered around $\Omega$
with variance $\Delta\omega^2$. The state $|\Psi\rangle_{en}$ can be
obtained from $|\Psi\rangle_G$ in the limit $\Delta\omega\to 0$.
Since $|\Omega\rangle_j$ has the same structure of
$|\Psi\rangle_{en}$, if one photon is lost in the $j$-th group all the
time of arrival information of such state must be discarded. Namely,
only the $g$ groups in which no photons have been lost can be still
employed for the positioning. In this case, using the state
$|\Psi\rangle_G$ in the ensemble average of Eq. (\ref{probmisn}) to
calculate the probability density of detecting all the $gK$ photons of
the $g$ groups at times $t_{j,l}$ is given by
\begin{eqnarray}
P_{gK}(t_{j,l})\propto \exp\left[-
{\left(\sum_{j=1}^g\sum_{l=1}^Kt_{j,l}\right)^2}/\left({2\Delta\tau_g^2}\right)
\right]
\;\label{biancaneve},
\end{eqnarray}
where $t_{j,l}$ is the time of arrival of the $l$-th photon in the
$j$-th group and\begin{eqnarray}
\Delta\tau_g=\frac{\sqrt{g}}{2\Delta\omega}
\sqrt{\frac{(G-g)\Delta\Omega^2+\Delta\omega^2}
{G\Delta\Omega^2+\Delta\omega^2}} 
\;\label{principeazzurro}.
\end{eqnarray}
Notice that Eq. (\ref{biancaneve}) and (\ref{principeazzurro}) for
$\Delta\omega\to 0$ and $G=g$ reproduce the result derived previously
in (\ref{classicp}) for Gaussian spectrum with $N=1$.
Eq. (\ref{biancaneve}) shows that even if $G-g$ groups are discarded
because they lost some photons, the remaining $g$ groups still retain
some entanglement. In fact, since the $|\Omega\rangle_j$ are not
orthogonal for $\Delta\omega>0$, the probability $P_{gK}(t_{j,l})$
does not factorize in parts depending on the single groups. The
proportionality constant in Eq. (\ref{biancaneve}) must be chosen so
that the integral of $P_{gK}(t_{j,l})$ over all the times gives the
probability that only $gK$ photons are detected, namely
\begin{eqnarray}
{\cal P}_g\equiv\left(\matrix{G\cr
g}\right)\frac
{\left(\eta^K\right)^g\left(1-\eta^K\right)^{G-g}}
{1-\left(1-\eta^K\right)^G}\;\label{settenani},
\end{eqnarray}
where $\eta^K$ is the probability that all the photons of a group
reach the detectors, and where, analogously as in
Sect. {\ref{s:pluto1}}, the term $1/{[1-\left(1-\eta^K\right)^G]}$ is
introduced to take into account the case (to be discarded) in which
all the $G$ groups have lost at least one photon.

If $g$ of the $G$ groups do not lose any photon, one may estimate the
mean time of arrival by calculating the mean value of
$\sum_{jl}t_{j,l}/(gK)$. The accuracy may be estimated by using the
probability (\ref{biancaneve}) obtaining \begin{eqnarray} &&\Delta
t=\frac 1{2K\Delta\omega}\label{pisolo}
\Big[\sum_{g=1}^G
\frac{(G-g)\Delta\Omega^2+\Delta\omega^2}
{g(G\Delta\Omega^2+\Delta\omega^2)}\;{\cal P}_g
\Big]^{\frac 12}
\;.
\end{eqnarray}
As before --see Eq. (\ref{varclass1})-- when $r\gg 1$ experimental
runs are performed, the accuracy $\Delta t(r)$ that can be achieved is
obtained from (\ref{pisolo}) by dividing $\Delta t$ by the square root
of the number of usable runs, namely $r[1-\left(1-\eta^K\right)^G]$.

\begin{figure}[hbt]
\begin{center}\epsfxsize=.6
\hsize\leavevmode\epsffile{figure7u.ps} 
\end{center}
\begin{center}
\epsfxsize=.8\hsize\leavevmode\epsffile{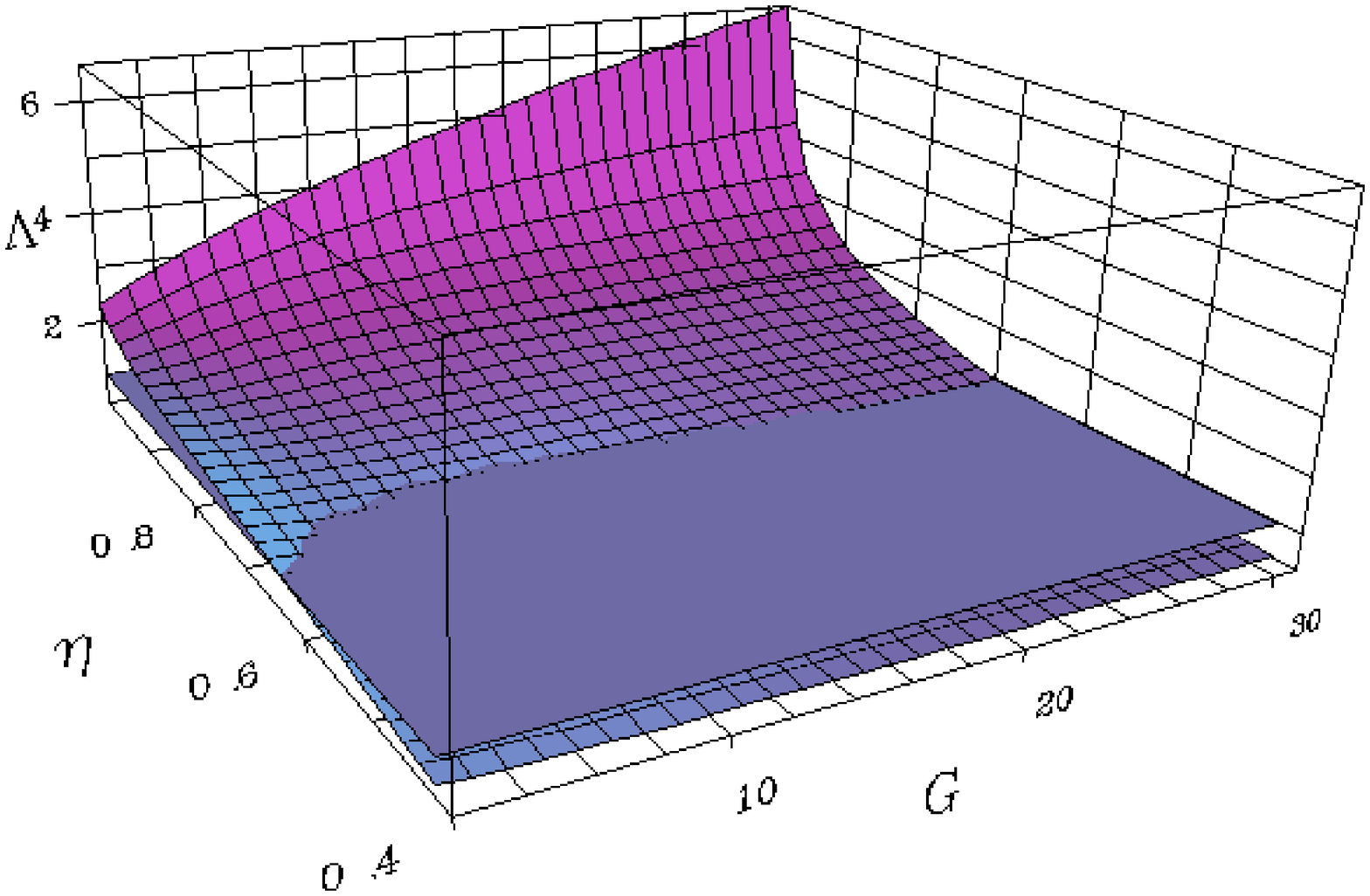} 
\end{center}
\caption{Robustness to loss of the state (\ref{omegag}). {\it Upper
graph:} The upper part of the graph shows for which values of the
quantum efficiency $\eta$ and of the total number of photons $M$ one
does better by using the state $|\Psi\rangle_G$ (with $K=4$ and
$\Delta\omega^2/\Delta\Omega^2=2$) as compared to the unentangled
state $|\Psi\rangle_{un}$. The dotted line is the same as in
Fig. {\ref{f:grafico}} and shows the region where it is better to use
maximally entangled states $|\Psi\rangle_{en}$ as compared to
unentangled ones $|\Psi\rangle_{un}$. {\it Lower graph:} The same
information as the previous graph is given plotted {\it vs.} the
number of photon groups $G$, but showing also the accuracy gain over
the unentangled case.}
\label{f:groupent}\end{figure}

\begin{figure}[hbt]
\begin{center}\epsfxsize=.6
\hsize\leavevmode\epsffile{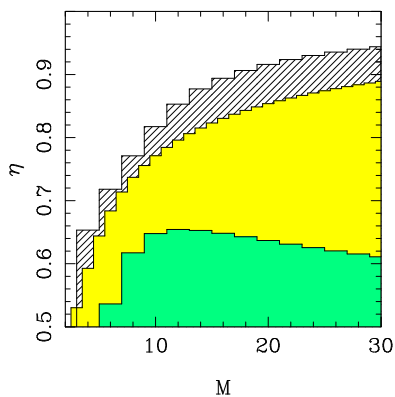} 
\end{center}
\caption{The upper white region is where the maximally entangled state
$|\Psi\rangle_{en}$ achieves a better accuracy than the
group-entangled state $|\Psi\rangle_G$ and than the unentangled state
$|\Psi\rangle_{un}$ (in brief: $en>G>un$). The striped region is where
$G>en>un$; the light grey region is where $G>un>en$, and the dark grey
region is where $un>G>en$. The parameters for this plot are $K=2$ and
$\Delta\omega^2/\Delta\Omega^2=2$. }
\label{f:groupmax}\end{figure}

In order to compare this result to what one would obtain in the
unentangled case or in the maximally entangled case, one must employ
the states $|\Psi\rangle_{en}$ and $|\Psi\rangle_{un}$ with the same
single photon spectral characteristics of the photons of
$|\Psi\rangle_G$. This can be achieved by using in $|\Psi\rangle_{en}$
and $|\Psi\rangle_{un}$ a Gaussian spectrum with variance
$\Delta\omega^2+\Delta\Omega^2$: namely,
$\Delta\tau=1/(2\sqrt{\Delta\omega^2+\Delta\Omega^2})$.  An example of
the comparison between the performance of $|\Psi\rangle_{un}$ and
$|\Psi\rangle_G$ when using such a coding scheme is given in
Fig. {\ref{f:groupent}}, where the group-entangled state
$|\Psi\rangle_G$ is shown to achieve a better accuracy than a
non-entangled state $|\Psi\rangle_{un}$. Notice that the accuracy
enhancement feature can be retained also for low quantum efficiency
even when a high number $M$ of particles is involved. A comparison
between the accuracy enhancement obtainable with the states
$|\Psi\rangle_{en}$, $|\Psi\rangle_{un}$ and $|\Psi\rangle_{G}$ is
shown in Fig. \ref{f:groupmax}.

\section{Conclusion}\label{s:concl}
In this paper, a scheme that employs entanglement and squeezing to
achieve a higher accuracy and cryptographic capabilities in position
measurement has been analyzed in detail. The positioning
quantum--cryptographic protocol described allows only trusted parties
(and no one else) to discover their relative positions.  The
sensitivity to the loss has been addressed by presenting a
quantitative analysis of different strategies to contrast it.
One finds that, even though the system is in principle very sensitive
to the loss of a single photon, there are many situations where it may
still be employed with an accuracy enhancement over the analogous
classical schemes. It has been shown that relaxing the requirements of
having maximally entangled states in frequency, one can achieve
greater resistance to losses.

An interesting feature, that has been analyzed elsewhere
{\cite{altropaper}}, is also present in our proposal. Namely, it is
possible to exploit the robustness of the frequency entanglement when
the pulses travel through dispersive media {\cite{kwiat}}. This may be
used to achieve positioning and clock synchronization of distant
parties without being affected by the intermediate dispersion that
would distort any timing signal the parties exchange.

\acknowledgments
This work was funded by the ARDA, NRO, and by ARO under a MURI
program.

\end{multicols}
\end{document}